\let\TeXyear\year
\documentclass[doublecol]{epl2} 
\let\setyear\year
\let\year\TeXyear

\usepackage{tikz,pgfplots}
\usetikzlibrary{arrows}
\usetikzlibrary{decorations.markings}
\usetikzlibrary{math}
\usetikzlibrary{positioning,calc}
\usetikzlibrary{backgrounds}
\usetikzlibrary{3d}
\usetikzlibrary{intersections, pgfplots.fillbetween}
\usetikzlibrary{patterns}
\usetikzlibrary{shapes.arrows}
\pgfplotsset{compat=1.16}

\setyear{2020}

\usepackage[utf8]{inputenc}
\usepackage[english]{babel}
\usepackage{amsmath}
\usepackage[T1]{fontenc}

\newcommand{\pot}{V}
\newcommand{\len}{L}
\newcommand{\f}{f}
\newcommand{\tm}{t}
\newcommand{\FPop}{\mathcal{L}}
\newcommand{\FPEV}{v}
\newcommand{\eval}{\lambda}
\newcommand{\hop}{H}
\newcommand{\ca}{a}
\newcommand{\potlr}{\tilde{\pot}}
\newcommand{\potr}{\pot_2}
\newcommand{\potir}{\pot_r}
\newcommand{\pp}{\varepsilon}
\newcommand{\sinco}{c}
\newcommand{\cosco}{d}
\newcommand{\C}{C}
\newcommand{\B}{B}

\newcommand{\dif}{\text{d}}
\newcommand\numberthis{\addtocounter{equation}{1}\tag{\theequation}}
\newcommand{\ronum}[1]{ \textup{\uppercase\expandafter{\romannumeral#1}}}

\tikzset{MyArrow/.style={single arrow, minimum width=2ex, minimum height=12ex, 
                         top color=red,bottom color=blue,rotate=-90,
                         single arrow head extend=1ex}}

\title{Anomalous relaxation from a non-equilibrium steady state: \\ An isothermal analog of the Mpemba effect}
\shorttitle{Anomalous relaxation: An isothermal analog of the Mpemba effect} %Insert here a short version of the title if it exceeds 70 characters

\author{Julius Degünther\inst{1} \and Udo Seifert\inst{1}}
\shortauthor{J. Degünther and U. Seifert}

\institute{                    
  \inst{1} II. Institut für Theoretische Physik, Universität Stuttgart, 70550 Stuttgart,
Germany
}

\abstract{The Mpemba effect denotes an anomalous relaxation phenomenon where a
system initially at a hot temperature cools faster than a system that
starts at a less elevated temperature. We introduce an isothermal analog
of this effect for a system prepared in a non-equilibrium steady state that
then relaxes towards equilibrium. Here, the driving strength, which
determines the initial non-equilibrium steady state, takes the role of
the temperature in the original version. As a paradigm, we consider a 
particle initially driven by a non-conservative force along a one-dimensional
periodic potential. We show that for an asymmetric potential relaxation
from a strongly driven initial state is faster than from a 
more weakly driven one at least for one of the two possible directions
of driving. These results are first obtained through perturbation theory
in the strength of the potential and then extended to potentials of 
arbitrary strength through topological arguments.}

\begin{document}

\maketitle

\section{Introduction}
For any initial preparation, a closed system will finally relax to an equilibrium state. Such relaxation processes feature a variety of intriguing and in parts counter-intuitive phenomena. One example is an asymmetry in heating and cooling processes where under certain circumstances the former is faster than the latter \cite{lapo20, van21, mani21, meib21}. Furthermore, introducing memory can give rise to anomalous relaxation processes characterized by power laws instead of exponentials \cite{metz99, bao19}. One of the most prominent examples for a surprising relaxation phenomenon is the Mpemba effect, which describes the observation that under certain conditions warmer water takes less time to freeze than colder one \cite{mpem69}. Since its discovery there has been extensive research concerning this particular phenomenon \cite{jeng06} with a multitude of possible origins such as different solute concentrations \cite{katz09}, supercooling \cite{auer95, moor11, brow11}, water hexamers \cite{jin15}, natural convection \cite{vynn15}, evaporation \cite{mira17}, breaking of energy equipartition \cite{gijo19} and hydrogen bonds \cite{zhan14, tao17}.

The Mpemba effect is not unique to water. While there is still some debate concerning water \cite{burr16}, this effect has recently been reported for a variety of different systems \cite{grea11,ahn16,lasa17,hu18,torr19,bait19,bisw20,yang20}. Furthermore, it can also be found in systems, which fall into the regime of stochastic thermodynamics \cite{lu17,klic19,kuma20,gal20,schw21,walk21,lin22}. These system are small enough for thermal fluctuations to be prominent in contrast to the macroscopic systems for which the effect was first documented. For these mesoscopic systems, a specific framework has been developed to quantify the Mpemba effect \cite{lu17,klic19}.

The wide range of systems, which show the Mpemba effect, suggests that its appearance may not be attributed to a specific property of the system but may be based on a more general mechanism. In fact, the question arises whether the Mpemba effect is specific to thermal relaxation or whether other kinds of relaxation processes can show a similar feature. 

\begin{figure}[t]
\begin{center}
\includegraphics[scale=1]{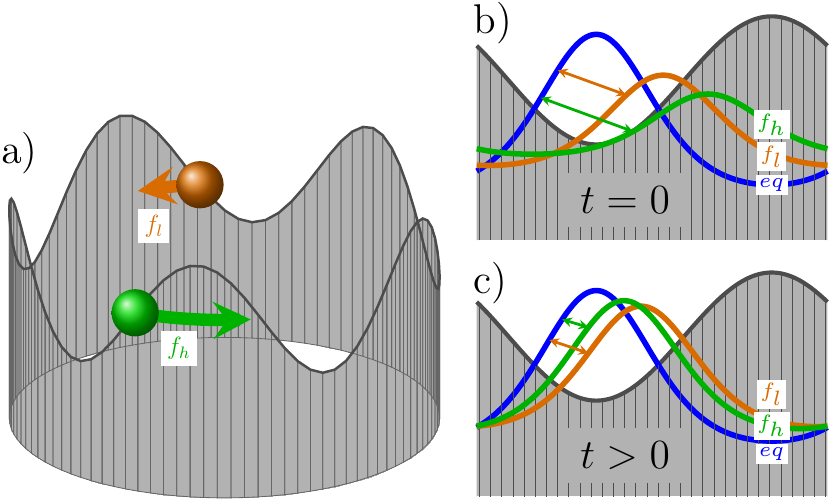}
  \caption{Illustration of the system and the isothermal analog of the Mpemba effect. a) The system consists of a particle on a ring.  The orange particle is driven by a weak force $f_l$ and the green one is driven by a strong force $f_h$. b) Sketch of a section of the initial non-equilibrium steady state distributions for both the green and the orange particle and the equilibrium distribution (blue). The distribution corresponding to $f_h$ is further from the equilibrium one. c) Sketch of a section of the distributions for the green and orange particle after some relaxation time and the equilibrium distribution. The distribution corresponding to $f_l$ is now further from the equilibrium one.}
  \label{fig:3}
\end{center}
\end{figure}

In this Letter, we address this question by considering relaxation from a non-equilibrium steady state into equilibrium. A system reaches a non-equilibrium steady state if it is subject to some form of time-independent driving. One of the major paradigms of stochastic thermodynamics is a Brownian particle driven by an external force. Under periodic boundary conditions, this system is arguably the simplest one reaching a non-equilibrium steady state, as illustrated in Figure \ref{fig:3}. Many theoretical predictions from stochastic thermodynamics \cite{seif12} have been experimentally verified using this system \cite{fauc95,blic07,gome09,mehl10}. Here, we will study the relaxation from its non-equilibrium steady state to equilibrium. We will demonstrate that an isothermal analog of the Mpemba effect is generic for this system. This Mpemba-like effect refers to the phenomenon that a system that is driven out of equilibrium by a stronger force relaxes faster than a system that is driven out of equilibrium by a weaker one, exemplarily illustrated by the two particles in Figure \ref{fig:3}. Intuitively, one would expect the contrary similar to the expectation that initially hot water takes longer to cool than less hot one.

\section{Setup and theory}
First, we define the system and its parameters. We consider a one-dimensional system with a continuous degree of freedom on a ring of length $\len$. The particle is subject to a non-conservative constant force $\f$ in addition to the potential landscape $\pot(x)$ with $\pot(x)=\pot(x+\len)$. The evolution of the probability to find the particle at position $x$ at time $\tm$ is governed by the Fokker-Planck equation. With energy given in units of $k_\text{B}T$ with $T$ the temperature and $k_\text{B}$ the Boltzmann constant and by rescaling the time coordinate $t=D\tilde{t}$, where $\tilde{t}$ is the original one and $D$ the diffusion constant, this equation reads 
\begin{equation} \label{eq:FPeq}
\partial_t p(x,\tm) = -\partial_x\left(F(x)-\partial_x\right)p(x,\tm)=\FPop p(x,\tm)
\end{equation}
with the Fokker-Planck operator $\FPop$ and the force $F(x)=\f-\partial_x\pot(x)$. Due to its periodicity, the potential can be expressed as a Fourier series
\begin{equation}
\pot(x)=\pp\sum_{k=1}^\infty\left[\sinco_k\sin\left(2\pi kx/L\right)+\cosco_k\cos\left(2\pi kx/L\right)\right],
\end{equation}
with the Fourier coefficients $\left\{\sinco_k,\cosco_k\right\}$ and a parameter $\pp$ that sets its overall scale.

In the absence of a driving force, this system has a unique equilibrium state defined by the potential $\pot(x)$. We calculate the speed of relaxation towards this equilibrium as a function of the driving force following what was done in \cite{lu17} for the thermal Mpemba effect. The solution of the Fokker-Planck equation \eqref{eq:FPeq} for a relaxation process towards equilibrium can formally be expressed by the series
\begin{equation} \label{eq:rel_sol}
p(x,t)=p^\text{eq}(x)+\sum_{n\geq2}\ca_n\FPEV_n(x) e^{\eval_n\tm},
\end{equation}
with the eigenvalues $\eval_n\leq0$ and corresponding real right eigenvectors $\FPEV(x)$ of $\FPop^\text{eq}$, coefficients $a_n$ determined by the initial distribution, and the equilibrium distribution $p^\text{eq}(x)=e^{-\pot(x)}/N^\text{eq}$ where $N^\text{eq}$ provides the normalization. $\FPop^\text{eq}$ denotes the Fokker-Planck operator without driving, \textit{i.e.}, for $\f=0$. It is advantageous to transform $\FPop^\text{eq}$ into a self-adjoint operator
\begin{equation}
\hop \equiv e^{\pot(x)/2}\FPop^\text{eq}e^{-\pot(x)/2}=\partial_x^2-\frac{(\partial_x\pot(x))^2}{4}+\frac{\partial_x^2\pot(x)}{2}
\end{equation}
with eigenvectors $\psi_n(x)$. The eigenvalues of $\hop$ and $\FPop^\text{eq}$ coincide and the eigenvectors are related via 
\begin{equation}
\psi_n(x)=e^{\pot(x)/2}\FPEV_n(x).
\end{equation}
This allows us to calculate the coefficients $a_n$  as
\begin{equation} \label{eq:a}
\ca_n = \int_0^L\dif xp(x,0)e^{\pot(x)/2}\psi_n(x),
\end{equation}
where $p(x,0)$ is the initial distribution. For sufficiently long time, the relaxation is dominated by the term in equation \eqref{eq:rel_sol} that corresponds to the second largest eigenvalue. We assume the eigenvalues are labeled in descending order $0=\eval_1>\eval_2\geq\eval_3\geq...$. Which of two given initial distributions relaxes faster towards the equilibrium state is therefore determined by the corresponding coefficients $\ca_2$. 

In our case, the initial distribution is the steady state $p(x,0)=p^{\text{ss}}(x;\f)$, which is determined by the driving strength $\f$. Therefore, we are interested in the monotonicity of the relaxation amplitude $\ca_2(\f)$. For simplicity, we omit the superscript and denote the steady state distribution by $p(x;\f)\equiv p^{\text{ss}}(x;f)$. Systems for which the relaxation amplitude $\ca_2(\f)$ is not monotonic show an isothermal analog of the Mpemba effect. Note that the steady state to $\f=0$ is the equilibrium distribution, which implies $\ca_2(0)=0$. 

\section{Perturbation theory}
Even for this minimal model, calculating $a_2(\f)$ analytically is not possible in general. Therefore, we first treat the above system with the additional assumption that the amplitude of the potential is small, \textit{i.e.}, $\pot(x)=\pp\potlr(x)$ with $\pp\ll 1$.  The assumption of a small potential allows us to employ perturbation theory. The ansatzes for the steady state distribution as well as the eigenvectors are
\begin{align} \label{eq:pert_ansatz}
p(x;\f) &= p^{(0)}(x;\f)+\pp p^{(1)}(x;\f)+\mathcal{O}(\pp^2) \\
\psi_n(x) &= \psi_n^{(0)}(x)+\pp\psi_n^{(1)}(x)+\mathcal{O}(\pp^2)
\end{align}
Inserting these into equation \eqref{eq:a} leads to the expansion
\begin{align*} \label{eq:aex}
&\ca_2(\f)=\int_0^L\dif x\Biggl\{p^{(0)}(x;\f)\psi_2^{(0)}(x) +\pp \Biggl[p^{(1)}(x;\f)\psi_2^{(0)}(x) \\
&+p^{(0)}(x;\f)\psi_2^{(0)}(x)\frac{\potlr(x)}{2}+p^{(0)}(x;\f)\psi_2^{(1)}(x)\Biggr]\Biggr\}+\mathcal{O}(\pp^2). \numberthis
\end{align*}
We perform the calculations up to linear order explicitly by inserting these expansions into the Fokker-Planck equation \eqref{eq:FPeq} and the eigenvalue problem 
\begin{equation} \label{eq:evp}
\hop\psi_n(x)=\eval_n\psi(x),
\end{equation}
respectively.

We first deal with the computation of the initial steady state. In leading order, the Fokker-Planck equation yields the equation for a particle on a ring without potential, which results in a uniform distribution $p^{(0)}(x;\f)=1/\len$ independently of the driving force $\f$. In linear order, the steady state Fokker-Planck equation reads 
\begin{equation} \label{eq:firstorderFP}
0 = \f p^{(1)}(x;\f)-\partial_x\potlr(x)p^{(0)}(x;\f)-\partial_xp^{(1)}(x;\f),
\end{equation}
where we use the normalization condition. The solution to equation \eqref{eq:firstorderFP} is given by
\begin{equation} \label{eq:psslr}
p^{(1)}(x;\f)=-\sum_{k=1}^\infty\frac{2\pi k}{\f^2\len^2+4\pi^2k^2}F_k(x)
\end{equation}
with
\begin{align*}
F_k(x)=&\left[\sinco_kk\frac{2\pi}{\len}+\cosco_k\f \right]\sin\left(k\frac{2\pi}{\len}x\right) \numberthis\\
+&\left[-\sinco_k\f+\cosco_k k\frac{2\pi}{\len}\right]\cos\left(k\frac{2\pi}{\len}x\right).
\end{align*}

\begin{figure*}[t]
\begin{center}
  \includegraphics[scale=0.34375]{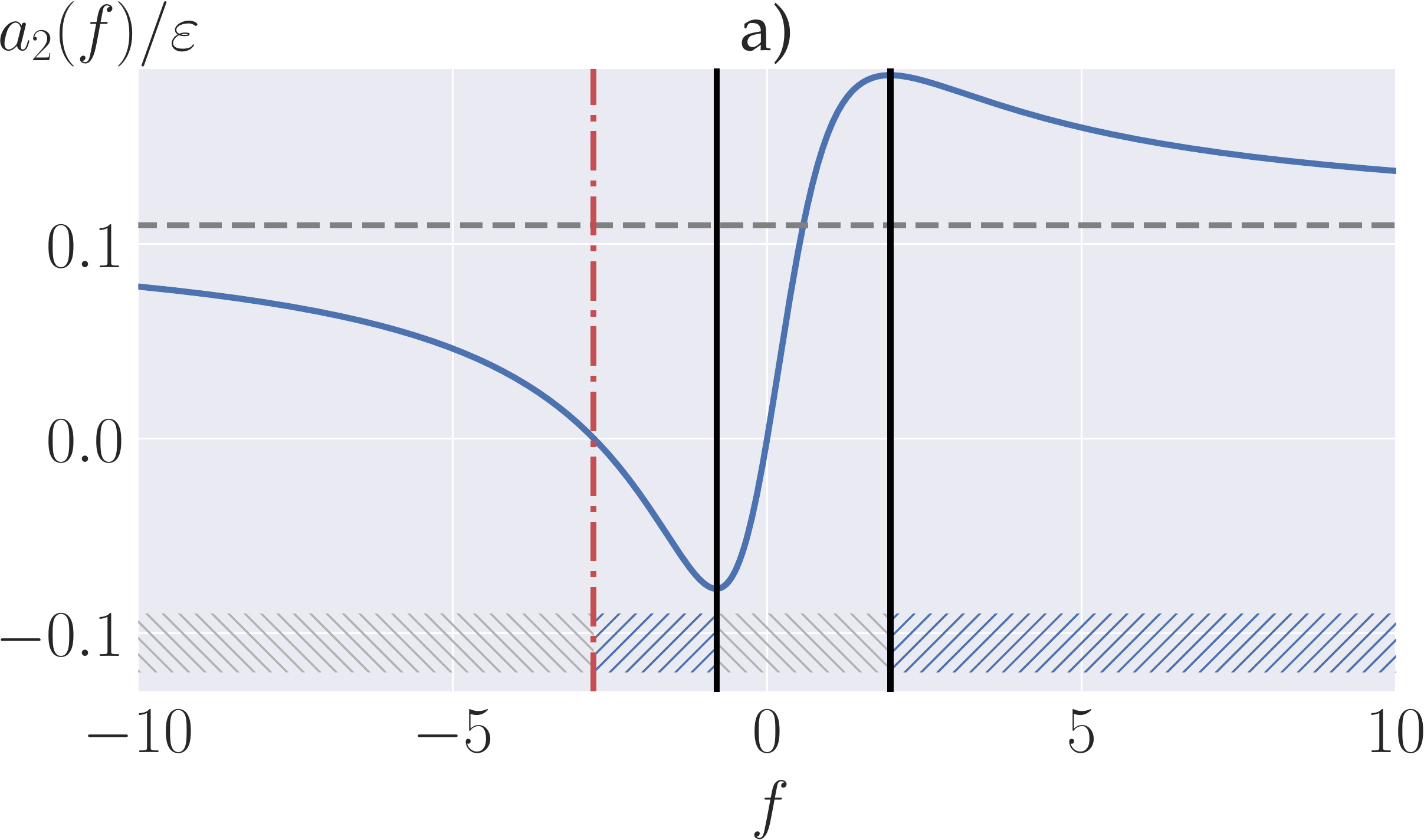}
  \hspace{1em}
  \includegraphics[scale=0.34375]{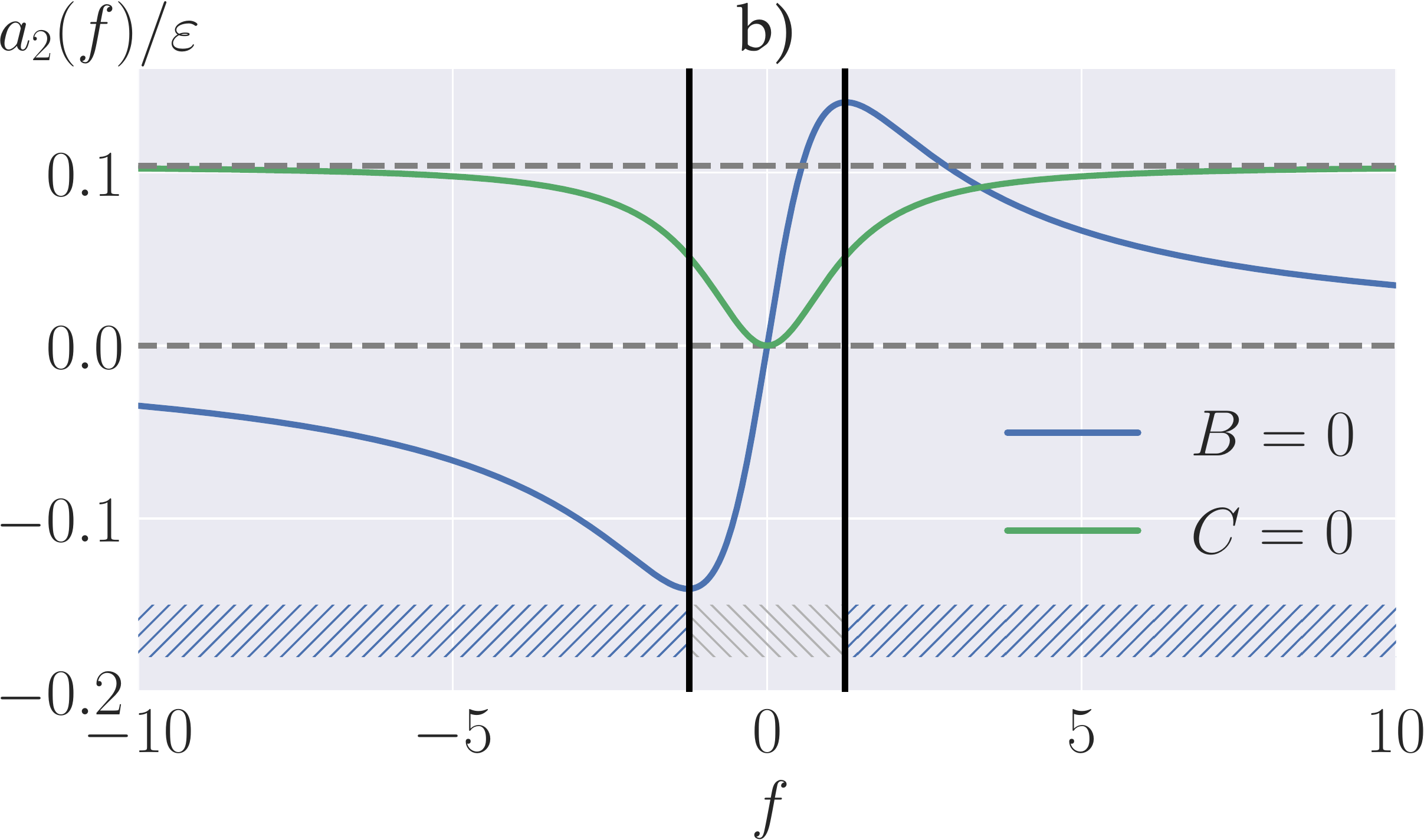}
  \caption{Relaxation amplitude $\ca_2(\f)$ for small potentials. a) Generic case: The black vertical lines mark the extrema of $\ca_2(\f)$. The grey bars (north west hatching) indicate areas without Mpemba effect. The blue bars (north east hatching) indicate areas where the Mpemba effect occurs, which means stronger driving leads to faster relaxation. The red dash-dotted vertical line marks the force $f_s$, which fulfills $\ca_2(\f_s)=0$ and, thus, shows the strong Mpemba effect. Parameters for this example $\{\sinco_1,\sinco_2,\sinco_3,\sinco_4\}=\{-0.82, -0.31, 0.57, -0.09\}$, $\{\cosco_1,\cosco_2,\cosco_3,\cosco_4\}=\{-0.17, -0.63, -0.25, 0.023\}$, $\len=5$ and $\pp=10^{-3}$. b) Symmetric potential; case\ronum{1}: $B=0$. The relaxation amplitude $\ca_2(\f)$ is antisymmetric with respect to $\f$ and does feature the Mpemba effect in both directions as marked by the blue bars (north east hatching). Parameters for this example $\sinco_i=0$, $\{\cosco_1,\cosco_2,\cosco_3,\cosco_4\}=\{0.45, 0.44, -0.98, 0.36\}$, $\len=5$ and $\pp=10^{-3}$. Case\ronum{2}: $C=0$. The relaxation amplitude $\ca_2(\f)$ is symmetric with respect to $\f$. This is the only case where no Mpemba effect can be observed at all. Parameters for this example $\sinco_i=0$, $\{\cosco_1,\cosco_2,\cosco_3,\cosco_4\}=\{-0.16, -0.06, -0.70, -0.03\}$, $\len=5$ and $\pp=10^{-3}$.}
  \label{fig:1}
\end{center}
\end{figure*}

Second, we need to solve the eigenvalue problem \eqref{eq:evp}. In leading order, this becomes the Schrödinger equation for a free particle with periodic boundary conditions. The ground state is $\psi_1^{(0)}(x)=1/\sqrt{\len}$ with the eigenvalue $\eval_1^{(0)}=0$ as expected. All further eigenvalues are twofold degenerate in lowest order. Thus, we need to apply degenerate perturbation theory. The first order corrections to the eigenvalues lifts the degeneracy and leads to the eigenvector corresponding to the second largest eigenvalue
\begin{equation} \label{eq:psi20}
\psi_2^{(0)}(x)=\left[\cosco_2+\sqrt{\sinco_2^2+\cosco_2^2}\right]\sin\left(\frac{2\pi}{L}x\right)-\sinco_2\cos\left(\frac{2\pi}{L}x\right)
\end{equation}
up to normalization. The last term in equation \eqref{eq:aex} requires the first order corrections to the eigenvectors. This leads to the result
\begin{equation} \label{eq:psi21}
\psi_2^{(1)}(x)=\frac{1}{\sqrt{8\len}}\frac{\sinco_1\cosco_2+\sinco_1\sqrt{\sinco_2^2+\cosco_2^2}-\cosco_1\sinco_2}{\sqrt{\left(\cosco_2+\sqrt{\sinco_2^2+\cosco_2^2}\right)^2+\sinco_2^2}}+G(x),
\end{equation}
where $G(x)$ is the part of $\psi_2^{(1)}(x)$ that vanishes upon integration over a full period $\len$. Since $p^{(0)}(x)$ is constant, only the part of $\psi_2^{(1)}(x)$ that does not vanish upon integration over one period matters for $\ca_2(\f)$. 

We finally have all the ingredients to evaluate the relaxation amplitude $\ca_2(\f)$. We insert equations \eqref{eq:psslr}, \eqref{eq:psi20} and \eqref{eq:psi21} into equation \eqref{eq:aex} and obtain up to linear order
\begin{equation} \label{eq:fullalr}
\ca_2(\f) = \frac{\pp}{\sqrt{2\len}}\left[\B-\frac{2\pi\len}{\f^2\len^2+4\pi^2}\left(\frac{2\pi}{\len}\B+f\C\right)\right]
\end{equation}
with the constants
\begin{align}
\B& = \frac{\sinco_1\cosco_2+\sinco_1\sqrt{\sinco_2^2+\cosco_2^2}-\cosco_1\sinco_2}{\sqrt{\left(\cosco_2+\sqrt{\sinco_2^2+\cosco_2^2}\right)^2+\sinco_2^2}}, \\
\C &= \frac{\cosco_1\cosco_2+\cosco_1\sqrt{\sinco_2^2+\cosco_2^2}+\sinco_1\sinco_2}{\sqrt{\left(\cosco_2+\sqrt{\sinco_2^2+\cosco_2^2}\right)^2+\sinco_2^2}}.
\end{align}
For now, we assume the generic case of $\B\neq0$ and $\C\neq0$ and deal with the special cases for which this is not true separately below. The zeroth order term of $\ca_2(\f)$ vanishes because it does not depend on $\f$ while it still has to fulfill the condition $\ca_2(0)=0$. Since we are interested in the monotonicity of $\ca_2(\f)$, we consider its derivative
\begin{equation} \label{eq:dfa2lr}
\partial_\f\ca_2(\f) =\frac{\pp2\pi/\sqrt{2\len^3}}{\left(\f^2+\frac{4\pi^2}{\len^2}\right)^2}\left[\C\f^2+\frac{4\pi}{\len}\B\f-\frac{4\pi^2}{\len^2}C\right].
\end{equation}
Any change in sign of $\partial_\f\ca_2(\f)$ implies a non-monotonic relaxation speed as a function of $\f$. Equation \eqref{eq:dfa2lr} shows that the derivative vanishes for $\f\to\pm\infty$ and for
\begin{equation} \label{eq:fex}
\f_{1,2}=-\frac{2\pi}{L}\left[\frac{\B}{\C}\pm\sqrt{1+\frac{\B^2}{\C^2}}\right].
\end{equation}
The asymptotic behavior for $\f\to\pm\infty$ is not related to a change in sign of $\partial_\f\ca_2(\f)$ and, therefore, does not imply a non-monotonicity. Thus, generically, there is one positive and one negative finite value for $\f$ at which $\partial_f\ca_2(\f)$ changes sign.  This leads to our first main result. For a generic small potential, $a_2(f)$ is not monotonic leading to an isothermal analog of the Mpemba effect. Here, generic means that the subspace of parameters that leads to a different behavior has a lower dimension than the full parameter space.

A special case occurs for $\ca_2=0$. In this case, the term corresponding to the second largest eigenvalue in \eqref{eq:rel_sol} vanishes and the third largest eigenvalue dominates the relaxation for sufficiently long times. This means that initial distributions for which $\ca_2=0$ relax exponentially faster than those with $\ca_2\neq0$. For thermal relaxation, this effect has previously been called strong Mpemba effect \cite{klic19}. We now want to analyze if an analog also occurs in our systems. The finite $\f_s$ for which $a_2(\f_s)=0$ obey
\begin{equation} \label{eq:a0}
\B\f_s^2-\frac{2\pi}{\len}\C\f_s=0.
\end{equation}
This equation has one non-trivial solution $\f_s\neq0$ for the general case of $\B\neq0$ and $\C\neq0$, which is either positive or negative. The sign of $\f$ determines the direction in which the particle is driven. This means that the strong Mpemba effect occurs for a generic potential, but only for one of the two possible signs of $\f$. This is in contrast to our findings concerning the normal Mpemba-like effect. While both are generic, the normal Mpemba-like effect occurs for both signs of $\f$ as the two possible solutions to equation \eqref{eq:fex} show.

Figure \ref{fig:1} a) summarizes the findings regarding the normal and strong Mpemba effect for the generic case. There is one particular value for $\f$ indicated by the red dash-dotted line for which the strong Mpemba effect occurs. Then there are two intervals in which the normal Mpemba effect occurs marked by the blue bars (north east hatching). One of these intervals is finite and bounded on one side by the force for which the strong Mpemba effect occurs. The other interval is not bounded. Here, the Mpemba effect is present for all forces beyond some critical force. 

In the above analysis of the normal and strong Mpemba effect we assumed generic potentials and disregarded special cases. More specifically, in both cases we assumed $\B\neq0$ and $\C\neq0$ and neglected the special cases in which this might not hold. We now examine these special cases, which behave differently. Note that only the Fourier coefficients $\left\{\sinco_1,\cosco_1,\sinco_2,\cosco_2\right\}$ of the first two terms of $\potlr(x)$ are relevant for $\ca_2(\f)$. For ease of description, we split the potential 
\begin{equation}
\pot(x)=\potr(x)+\potir(x)
\end{equation}
into the part $\potr(x)$ that is relevant for the Mpemba effect and into the irrelevant part $\potir(x)$. As it turns out, either of the conditions $\B=0$ and $\C=0$ implies that the potential $\potr(x)$ is symmetric. Which of the two conditions is met in the case of a symmetric potential depends on its exact form. By symmetric potentials we denote any potential with a symmetry $\potr(x_0+x)=\potr(x_0-x)$ with $x_0\in\left[0,L\right]$, which accounts for the translation invariance of the system. In Figure \ref{fig:1} b) the two possible cases $\B=0$ and $\C=0$ are sketched. As is obvious from equation \eqref{eq:fullalr}, the relaxation amplitude $\ca_2(\f)$ is antisymmetric for $\B=0$. Thus, there still is the normal Mpemba effect but there is no strong Mpemba effect. For $\C=0$  the relaxation amplitude $\ca_2(\f)$ is symmetric and the system shows neither the normal nor the strong Mpemba effect. This is the only case where no Mpemba effect occurs at all. 

The fact that symmetric potentials pose an exception is in agreement with the above generic findings about the strong Mpemba effect. The symmetry $\potr(x_0+x)=\potr(x_0-x)$ makes both directions of driving equivalent whereas we have derived that only one sign can show the strong Mpemba effect. The above generic results are, thus, valid for any potential that is not symmetric.
\begin{figure}[t]
\begin{center}
\includegraphics[scale=1]{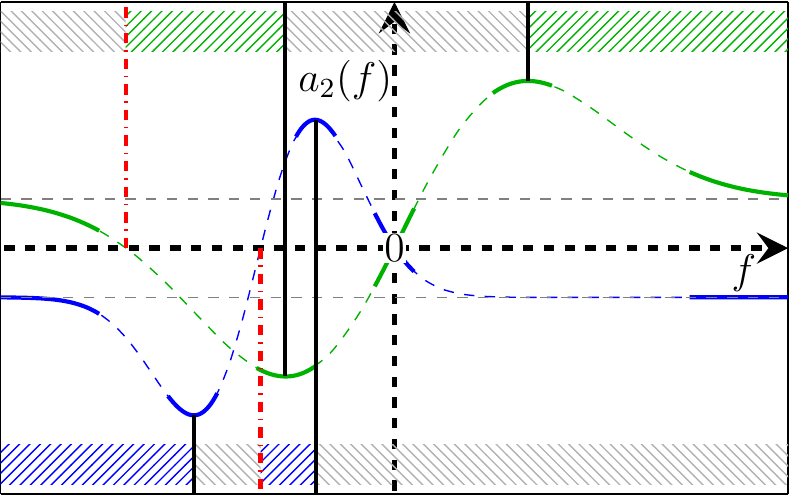}
  \caption{Qualitative illustration of the two alternatives for the relaxation amplitude $\ca_2(\f)$ for a generic potential with arbitrary strength. The solid parts are known features of the curves. The dashed parts are unknown except for that they are continuous. The colored bars (north east hatching) indicate where the Mpemba effect occurs. The grey bars (north west hatching) indicate areas of normal relaxation. The red dash-dotted lines mark the strong Mpemba effect. The main difference between the two cases is that the green curve has one extremum for each sign of $\f$, whereas the blue curve has two extrema for one sign and none for the other.}
  \label{fig:2}
\end{center}
\end{figure}	

\section{Beyond small potentials}
We will now argue that these results obtained perturbatively for a small potential hold true even for an arbitrarily strong potential. We show this by a topological argument using the asymptotic behavior of $\ca_2(\f)$ for strong driving and $\ca_2(0)=0$. Note that the force dependence of $\ca_2(\f)$ stems from the steady state while the eigenvectors are a property of the equilibrium and, therefore, are independent of $\f$. For large $|\f|$, the steady state solution of the Fokker-Planck equation can be expanded in orders of $1/\f$, which leads to
\begin{equation}
p(x;\f)=\frac{1}{L}\left(1+\frac{1}{f}\partial_x\pot(x)\right)+\mathcal{O}\left(\frac{1}{f^2}\right).
\end{equation}
Inserting this expansion into $\ca_2(\f)$ given by equation \eqref{eq:a} yields
\begin{align}\label{eq:loa}
\ca_2(\f) &= \int_0^\len\dif x\frac{1}{L}e^{\pot(x)/2}\psi_2(x)+\mathcal{O}\left(\frac{1}{f}\right)
\end{align} 
and
\begin{align} \label{eq:loda}
\partial_f\ca_2(\f) &= -\frac{1}{f^2}\int_0^\len\dif x\frac{1}{L}\partial_x\pot(x)e^{\pot(x)/2}\psi_2(x)+\mathcal{O}\left(\frac{1}{f^3}\right),
\end{align}
in leading order. Notably, both terms are independent of the sign of $\f$ in leading order, which implies
\begin{align}
\lim_{\f\to\infty}\ca_2(\f)&=\lim_{\f\to-\infty}\ca_2(\f) \\
\lim_{\f\to\infty}\partial_f\ca_2(\f)&=\lim_{\f\to-\infty}\partial_f\ca_2(\f).
\end{align}
Additionally, we use the property that $\ca_2(\f)$ vanishes for $\f=0$. This is an immediate consequence of the fact that the steady state defined by $\f=0$ is the equilibrium distribution. With this knowledge we can infer crucial properties of $\ca_2(\f)$. The relaxation amplitude $\ca_2(\f)$ falls in one of two classes illustrated in Figure \ref{fig:2}. The derivative $\partial_f\ca_2(\f)$ generically has at least two changes of sign because it is identical for $\f\to\infty$ and $\f\to-\infty$. These can either both occur for the same sign of $\f$ as the blue curve indicates or they can lie on different half-axes as for the green one. Note that with this topological reasoning we assume nothing specific about the dashed parts in Figure \ref{fig:2} except that $\ca_2(\f)$ is continuous. Thus, $\ca_2(\f)$ can have more extrema and zeros than the ones indicated; however, it can not have less. This means that even for an arbitrary potential we generically predict the Mpemba effect for at least one of the two signs of $\f$. Regarding the strong Mpemba effect, we conclude that $\ca_2(\f)$ generically has at least one zero besides the trivial one at $\f=0$. Thus, the strong Mpemba effect is guaranteed for at least one sign of $\f$, similar to the case with a small potential. Exceptions from this generic behavior arise if any of the ordinarily leading orders in $\ca_2(\f)$ or $\partial_f\ca_2(\f)$ in equations \eqref{eq:loa} and \eqref{eq:loda} vanish. A further exceptional case occurs when the derivative $\partial_f\ca_2(\f)$ vanishes at $\f=0$, which allows $\ca_2(\f)$ not to change its sign. For these special cases the above implications do not necessarily hold. As these results demonstrate, many of the findings in the limit of a small potential thus carry over to the general case. 

\section{Conclusion}
We have introduced an isothermal analog of the Mpemba effect. The system is initially prepared in a non-equilibrium steady state and relaxes towards equilibrium. The driving force, which ultimately determines the steady state, takes the role of the initial temperature in the classical Mpemba effect. If a stronger initial driving force leads to faster relaxation, the isothermal analog of the Mpemba effect arises. 

We have shown that this Mpemba-like effect is generic. We explicitly calculate all relevant quantities in the limit of an arbitrary but small potential establishing both the normal and the strong Mpemba effect. The only exceptions to these generic findings arise when the potential is symmetric. Through topological arguments we have extended these results to the case of an arbitrary potential for at least one sign of the non-conservative driving force. 

In this Letter, we have considered a continuous dynamics governed by a 
one-dimensional Fokker-Planck equation. For such a dynamics in two
or three dimensions, or even for interacting Langevin particles, one 
might expect a similar phenomenology which deserves to be investigated in
detail. Moreover,
it remains to be seen whether similar results hold for a Markovian dynamics
on a discrete set of states. To address this question, a first step would
be to study the behavior for a unicyclic system. While the occurrence of
the present analog of the Mpemba effect might still be generic, we suspect that 
there could be qualitative differences to the results obtained within the 
continuous model. The main reason is that the continuous system behaves similarly in the limits of strong driving regardless of the sign of the driving force. This symmetry
can generally not be expected in discrete systems that are, \textit{e.g.}, 
driven by a non-equilibrium chemical reaction. Finally, it would be desirable
to search for an experimental realization, which should be feasible for
the paradigmatic driven particle.

\bibliography{Mpemba.bib}
%\bibliography{./Bibliography/refs.bib}
\bibliographystyle{eplbib}

\end{document}